\documentclass[conference]{IEEEtran}
\usepackage{graphicx}
\usepackage{amsmath}
\usepackage{amsfonts}
\usepackage{amssymb}
\usepackage{amsthm}
\usepackage{float}
\usepackage{stfloats}%to allow double column floats at the bottom of pages
\usepackage{topcapt}
\usepackage{hyperref}

\newcommand{\hv}{{\bf h}}

\newcommand{\sv}{{\bf s}}

\newcommand{\wv}{{\bf w}}

\newcommand{\yv}{{\bf y}}

\newcommand{\zerov}{{\bf 0}}

\newcommand{\Am}{{\bf A}}
\newcommand{\Bm}{{\bf B}}
\newcommand{\Cm}{{\bf C}}

\newcommand{\Hm}{{\bf H}}
\newcommand{\Id}{{\bf I}}

\newcommand{\Qm}{{\bf Q}}
\newcommand{\Rm}{{\bf R}}

\newcommand{\Um}{{\bf U}}
\newcommand{\Wm}{{\bf W}}

\newcommand{\Xm}{{\bf X}}
\newcommand{\Ym}{{\bf Y}}

\newcommand{\Ac}{{\cal A}}
\newcommand{\Bc}{{\cal B}}
\newcommand{\Cc}{{\cal C}}

\newcommand{\Gc}{{\cal G}}

\newcommand{\Nc}{{\cal N}}
\newcommand{\Oc}{{\cal O}}

\newcommand{\Hcb}{\pmb{\cal H}}

\newtheorem{mydef}{Definition}
\newtheorem{myprop}{Proposition}

\allowdisplaybreaks
\begin{document}
\title{A New Family of Low-Complexity Decodable STBCs for Four Transmit Antennas}
\author{\IEEEauthorblockN{Amr Ismail, Jocelyn Fiorina, and Hikmet Sari}
\IEEEauthorblockA{Telecommunications Department, SUPELEC, F-91192 Gif-sur-Yvette,France\\
Email:$\lbrace$amr.ismail, jocelyn.fiorina, and hikmet.sari$\rbrace$@supelec.fr\\}}

\maketitle
\begin{abstract}
In this paper we propose a new construction method for rate-1 Fast-Group-Decodable (FGD) Space-Time-Block Codes (STBC)s for $2^a$ transmit antennas. We focus on the case of $a=2$ and we show that the new FGD rate-1 code has the lowest worst-case decoding complexity among existing comparable STBCs. The coding gain of the new rate-1 code is then optimized through constellation stretching and proved to be constant irrespective of the underlying QAM constellation prior to normalization. In a second step, we propose a new rate-2 STBC that multiplexes two of our rate-1 codes by the means of a unitary matrix. A compromise between rate and complexity is then obtained through puncturing our rate-2 code giving rise to a new rate-3/2 code. The proposed codes are compared to existing codes in the literature and simulation results show that our rate-3/2 code has a lower average decoding complexity while our rate-2 code maintains its lower average decoding complexity in the low SNR region at the expense of a small performance loss.
\end{abstract}

\section{Introduction}
The need for low-complexity decodable STBCs is inevitable in the case of high-rate communications over MIMO systems employing a number of transmit antennas higher than two. The decoding complexity may be evaluated by different measures, namely the worst-case decoding complexity measure and the average decoding complexity measure. The worst-case decoding complexity is defined as the minimum number of times an exhaustive search decoder has to compute the Maximum Likelihood (ML) metric to optimally estimate the transmitted symbols codeword \cite{FAST,GLOB10}, or equivalently the number of leaf nodes in a search tree if a sphere decoder is employed, whereas the average decoding complexity measure may be numerically evaluated as the average number of visited nodes by a sphere decoder in order to optimally estimate the transmitted symbols codeword \cite{MIMO_VLSI}. Arguably, the first proposed low-complexity rate-1 code for the case of four transmit antennas is the Quasi-Orthogonal (QO)STBC originally proposed by H. Jafarkhani \cite{QOSTBC} and later optimized through constellation rotation to provide full diversity \cite{FULL_DIVERSITY_QOD,OPT_QORTH_ROT}. The QOSTBC partially relaxes the orthogonality conditions by allowing two complex symbols to be jointly detected. Subsequently, rate-1, full-diversity QOSTBCs were proposed for an arbitrary number of transmit antennas that subsume the original QOSTBC as a special case \cite{GEN_QOSTBC}. In this general framework, the \textit{quasi-orthgonality} stands for decoupling the transmitted symbols into two groups of the same size. However, STBCs with lower decoding complexity may be obtained through the concept of multi-group decodability laid by the S. Karmakar \textit{et al.} in \cite{MULTI_SYMs,MULT_GR}. Indeed, the multi-group decodability generalizes the quasi-orthogonality by allowing more than two groups of symbols to be decoupled not necessarily with the same size. 

However, due to the strict rate limitation imposed by the multi-group decodability, another family of STBCs namely Fast Decodable (FD) STBCs \cite{FAST} has been proposed. These codes are conditionally multi-group decodable thus enabling the use of the conditional detection technique \cite{2TX_MPLX_ORTH} which in turn significantly reduces the overall decoding complexity. Recently, STBCs that combine the multi-group decodability and the fast decodability namely the Fast-Group Decodable (FGD) codes have been proposed \cite{FGD}. These codes are multi-group-group decodable such that each group of symbols is fast decodable. The contributions of this paper are summarized in the following:
\begin{itemize}
\item We propose a novel systematic construction of rate-1 FGD STBCs for $2^a$ transmit antennas. The rate-1 FGD code for a number of transmit antennas that is not a power of two is obtained by removing the appropriate number of columns from the rate-1 FGD code corresponding to the nearest greater number of antennas that is a power of two (e.g. the rate-1 FGD STBC for three transmit antennas is obtained by removing a single column from the four transmit antennas rate-1 FGD STBC). 
\item We apply our new construction method to the case of four transmit antennas and show that the resulting new 4$\times$4 rate-1 code can be decoded at half the worst-case decoding complexity of the best known rate-1 STBC.
\item The coding gain of the new 4$\times$4 rate-1 code is optimized through constellation stretching \cite{CONST_STRETCHING} and the NonVanishing Determinant (NVD) property \cite{NVD} is proven to be achieved by properly choosing the stretching factor. 
\item We propose a new rate-2 STBC through multiplexing two of the new rate-1 codes by the means of a unitary matrix and numerical optimization. We then propose a significant reduction of the worst-decoding complexity at the expense of a rate loss by puncturing the rate-2 code to obtain a new rate-3/2 code. 
\end{itemize}
We compare the proposed codes to existing STBCs in the literature and found through numerical simulations that our rate-3/2 code has a significantly lower average decoding complexity while our rate-2 code is decoded with a lower average decoding complexity at low SNR region. Performance simulations show that this reduction in the decoding complexity comes at the expense of a small performance loss.

The rest of the paper is organized as follows: The system model is defined and the families of low-complexity STBCs are outlined in Section II. In Section III we propose our scheme for the rate-1 FGD codes construction for the case of $2^a$ transmit antennas, and then the FGD code construction method is applied to the case of four transmit antennas giving rise to a new 4$\times$4 rate-1 STBC. In Section IV, the rate of the proposed code is increased through multiplexing and numerical optimization. Numerical results are provided in Section V, and we conclude the paper in Section VI. 
\subsection*{Notations:}
Hereafter, small letters, bold small letters and bold capital letters will designate scalars, vectors and matrices, respectively. If $\Am$ is a matrix, then $\Am^H$, and $\Am^T$ denote the hermitian and the transpose of $\Am$, respectively. We define the $\textit{vec}(.)$ as the operator which, when applied to  a $m \times n$ matrix, transforms it  into a $mn\times 1$ vector by simply concatenating vertically the columns of the corresponding matrix. The $\otimes$ operator is the Kronecker product and the $\textit{sign}(.)$ operator returns 1 if its scalar input is $\geq 0$ and -1 otherwise. The $\textit{round}(.)$ operator rounds its argument to the nearest integer. The $\tilde{(.)}$ operator concatenates vertically the real and imaginary parts of its argument. $\left(a\right)_n$ means $a$ modulo $n$.
%************************************************************************************************************%
\begin{table*}[b!]
\centering
\topcaption{Different cases for $\Ac$ and $\Bc$}
\tiny
\begin{tabular}{|c|c|c|}
\hline
$a$&$\Ac$&$\Bc$\\
\hline
$4n$&$\left\lbrace j\prod^{2a+1}_{i=a+1}\Rm_i\right\rbrace \overset{a-2}{\underset{m=1}{\cup}} \left\lbrace j^{\delta_{\Ac}(m)}\prod^{2a+1}_{i=a+1}\Rm_i\prod^{m}_{i=1}\Rm_{k_i}:1\leq k_1<\ldots<k_m\leq a \right\rbrace$&
$\left\lbrace j\prod^{a}_{i=1}\Rm_i\right\rbrace \overset{a-2}{\underset{m=2,4}{\cup}}\left\lbrace j^{\delta_{\Bc}(m)}\prod^{a}_{i=1}\Rm_i\prod^{m}_{i=1}\Rm_{k_i}:a+1\leq k_1<\ldots<k_m\leq 2a+1 \right\rbrace$\\
\hline
$4n+1$&$\left\lbrace \prod^{2a+1}_{i=a+1}\Rm_i\right\rbrace \overset{a-2}{\underset{m=1}{\cup}} \left\lbrace j^{\delta_{\Ac}(m)}\prod^{2a+1}_{i=a+1}\Rm_i\prod^{m}_{i=1}\Rm_{k_i}:1\leq k_1<\ldots<k_m\leq a \right\rbrace$&
$\overset{a-2}{\underset{m=1,3}{\cup}}\left\lbrace j^{\delta_{\Bc}
(m)}\prod^{a}_{i=1}\Rm_i\prod^{m}_{i=1}\Rm_{k_i}:a+1\leq k_1<\ldots<k_m\leq 2a+1 \right\rbrace$\\
\hline
$4n+2$&$\left\lbrace \prod^{2a+1}_{i=a+1}\Rm_i\right\rbrace \overset{a-2}{\underset{m=1}{\cup}}\left\lbrace  j^{\delta_{\Ac}(m)}\prod^{2a+1}_{i=a+1}\Rm_i\prod^{m}_{i=1}\Rm_{k_i}:1\leq k_1<\ldots<k_m\leq a \right\rbrace$&
$\left\lbrace \prod^{a}_{i=1}\Rm_i\right\rbrace \overset{a-2}{\underset{m=2,4}{\cup}}\left\lbrace j^{\delta_{\Bc}(m)}\prod^{a}_{i=1}\Rm_i\prod^{m}_{i=1}\Rm_{k_i}:a+1\leq k_1<\ldots<k_m\leq 2a+1 \right\rbrace$\\
\hline
$4n+3$&$\left\lbrace j\prod^{2a+1}_{i=a+1}\Rm_i\right\rbrace \overset{a-2}{\underset{m=1}{\cup}}\left\lbrace  j^{\delta_{\Ac}(m)}\prod^{2a+1}_{i=a+1}\Rm_i\prod^{m}_{i=1}\Rm_{k_i}:1\leq k_1<\ldots<k_m\leq a \right\rbrace$&
$\overset{a-2}{\underset{m=1,3}{\cup}}\left\lbrace j^{\delta_{\Bc}(m)}\prod^{a}_{i=1}\Rm_i\prod^{m}_{i=1}\Rm_{k_i}:a+1\leq k_1<\ldots<k_m\leq 2a+1 \right\rbrace$\\
\hline
\end{tabular}
\label{casesI}
\end{table*}
%******************************************************************************************************************%
\section{Preliminaries}
We define the MIMO channel input-output relation as: 
\begin{equation}
\underset{T\times N_r}{\Ym} =\underset{T\times N_t}{\Xm} \underset{N_t\times N_r}{\Hm} +\underset{T\times N_r}{\Wm}
\label{model}  
\end{equation} 
where $T$ is the number of channel uses, $N_r$ is the number of receive antennas, $N_t$ is the number of transmit antennas, $\Ym$ is the received signal matrix, $\Xm$ is the code matrix, $\Hm$ is the channel matrix with entries $h_{kl} \sim \Cc \Nc(0,1)$, and $\Wm$ is the noise matrix with entries $w_{ij} \sim \Cc \Nc(0,N_{0} )$. In the case of Linear Dispersion (LD) codes \cite{LDSTBC}, a STBC that encodes $2K$ real symbols is expressed as a linear combination of the transmitted symbols as:
\begin{equation}
\Xm=\sum^{2K}_{k=1} \Am_kx_k
\label{LD}
\end{equation}
with $x_k\in\mathbb{R}$ and the $\Am_k, k=1,...,2K$ are $T \times N_t$ complex matrices called dispersion or weight matrices that are required to be linearly independent over $\mathbb{R}$. The MIMO channel model can then be expressed in a useful manner by using \eqref{LD} as:
\begin{equation}
\Ym=\sum^{2K}_{k=1}\left(\Am_k\Hm\right)x_k+\Wm.
\end{equation} 
Applying the $\textit{vec}(.)$ operator to the above equation we obtain:
\begin{equation}
\textit{vec}(\Ym)=\sum^{2K}_{k=1}\left(\Id_{N_r}\otimes\Am_k\right)\textit{vec}\left(\Hm\right)x_k+\textit{vec}(\Wm).
\label{vec}
\end{equation}
where $\Id_{N_r}$ is the $N_r\times N_r$ identity matrix. If $\yv_i$, $\hv_i$ and $\wv_i$ designate the $i$'th column of the received signal matrix $\Ym$, the channel matrix $\Hm$ and the noise matrix $\Wm$ respectively, then equation \eqref{vec} can be written in  matrix form as:
\small
\begin{equation}
\underbrace{\begin{bmatrix}\yv_1\\\vdots\\\yv_{N_r}\end{bmatrix}}_{\yv}=\underbrace{\begin{bmatrix}\Am_{1}\hv_1&\dots&\Am_{2K}\hv_1\\\vdots&\vdots&\vdots\\\Am_1\hv_{N_r}&\dots&\Am_{2K}\hv_{N_r}\end{bmatrix}}_{\Hcb}\underbrace{\begin{bmatrix}x_1\\\vdots\\x_{2K}\end{bmatrix}}_{\sv}
+\underbrace{\begin{bmatrix}\wv_1\\\vdots\\\wv_{N_r}\end{bmatrix}}_{\wv}.
\end{equation}
\normalsize
Thus we have:
\begin{equation}
\yv=\Hcb\sv+\wv
\label{model2}
\end{equation}
A real system of equations is be obtained by applying\\ the $\tilde{\left(.\right)}$ operator to the \eqref{model2}:
\begin{equation}
\tilde{\yv}=\tilde{\Hcb}\sv+\tilde{\wv}
\label{real_model}
\end{equation}
where $\yv,\wv \in \mathbb{R}^{2N_rT\times 1}$, and $\tilde{\Hcb}\in\mathbb{R}^{2N_rT\times 2K}$.  
Assuming that $N_rT \geq K$, the QR decomposition of $\tilde{\Hcb}$ yields: 
\begin{equation}
\tilde{\Hcb}=\begin{bmatrix}\Qm_1 & \Qm_2\end{bmatrix} \begin{bmatrix}\Rm \\ \zerov \end{bmatrix}
\end{equation}
where $\Qm_1\in \mathbb{R}^{2N_rT\times 2K}$,$\Qm_2\in\mathbb{R}^{2N_rT \times(2N_rT-2K)}$,  $\Qm_i^T\Qm_i=\Id,\ i=1,2$, $\Rm$ is a ${2K \times 2K}$ real upper triangular matrix and $\zerov$ is a 
$(2N_rT-2K)\times 2K$ null matrix.
Accordingly, the ML estimate may be expressed as:
\begin{equation}
\sv^{\text{ML}}=\text{arg}\ \underset{\sv \in \Cc}{\text{min}}\Vert \tilde{\yv}-\Qm_1\Rm\sv \Vert^2
\end{equation}
where $\Cc$ is the vector space spanned by information vector $\sv$. Noting that multiplying a column vector by a unitary matrix does not alter its norm, the above reduces to:
\begin{equation}
\sv^{\text{ML}}=\text{arg}\ \underset{\sv \in \Cc}{\text{min}}\Vert \yv'-\Rm\sv \Vert^2
\label{R}
\end{equation}
where $\yv'=\Qm_1^T\tilde{\yv}$.

In the following, we will briefly review the known families of low-complexity STBCs and the structures of their corresponding $\Rm$ matrices that enable a simplified ML detection.
\subsection{Multi-group decodable codes}
Multi-group decodable STBCs are designed to significantly reduce the worst-case decoding
complexity by allowing separate detection of disjoint groups of symbols without any loss of
performance. This is achieved iff the ML metric can be expressed as a sum of terms depending
on disjoint groups of symbols.
\begin{mydef}A STBC code that encodes $2K$ real symbols is said to be $g$-group decodable if its weight matrices are such that \cite{MULT_GR,MULTI_SYMs}: 
\begin{equation}
\begin{split}
\Am_k^{H}\Am_l+\Am_l^{H}\Am_k=\pmb{0},\ \forall \Am_k \in \Gc_i,\ \Am_l \in \Gc_j,\\
1\leq i\neq j\leq g,\ \vert \Gc_i \vert=n_i,\sum^{g}_{i=1}n_i=2K.
\end{split}
\label{g-group}
\end{equation}
where $\Gc_i$ is the set of weight matrices associated to the $i$'th group of symbols.
\end{mydef}
\noindent For instance if a STBC that encodes $2K$ real symbols is $g$-group decodable, its worst-case decoding complexity order can be reduced from $M^K$ to $\sum^{g}_{i=1}\sqrt{M}^{n_i}$ with $M$ being the size of the used square QAM constellation. The worst-case decoding complexity order can be further reduced to $\sum^{g}_{i=1}\sqrt{M}^{n_i-1}$ if the conditional detection with hard slicer is employed. In the special case of orthogonal STBCs, the worst-case decoding complexity is $\Oc(1)$ as the PAM slicers need only a fixed number of arithmetic operations irrespectively of the square QAM constellation size. 

\subsection{Fast decodable codes}
A STBC is said to be fast decodable if it is conditionally multi-group decodable. 
\begin{mydef}A STBC that encodes $2K$ real symbols is said to be FD if its weight matrices are such that:
\begin{equation}
\begin{split}
\Am_k^{H}\Am_l+\Am_l^{H}\Am_k=\pmb{0},\ \forall \Am_k \in \Gc_i,\ \Am_l \in \Gc_j,\\
1\leq i\neq j\leq g,\ \vert \Gc_i\vert=n_i,\sum^{g}_{i=1}n_i=k<2K.
\end{split}
\end{equation}
where $\Gc_i$ is the set of weight matrices associated to the $i$'th group of symbols.
\end{mydef}
\noindent In this case the conditional detection may be used to significantly reduce the worst-case decoding complexity. The first step consists of evaluating the ML estimate of $\left(x_1,\ldots,x_k\right)$ conditioned on a given value of the rest of the symbols $\left(\hat{x}_{k+1},\ldots,\hat{x}_{2K}\right)$ that we may note by $\left(x^{\text{ML}}_1,\ldots,x^{\text{ML}}_k\vert\hat{x}_{k+1},\ldots,\hat{x}_{2K}\right)$. In the second step, the receiver will have to minimize the ML metric only over all the possible values of $\left(x_{k+1},\ldots,x_{2K}\right)$. For instance, if a STBC that encodes $2K$ real symbols is FD, its corresponding worst-case decoding complexity order for square QAM constellations is reduced from $M^K$ to $\sqrt{M}^{2K-k}\times \sum^{g}_{i=1}\sqrt{M}^{n_i-1}$. If the FD code is in fact conditionally orthogonal, the worst-case decoding complexity order is reduced to $\sqrt{M}^{2K-k}$.

\subsection{Fast group decodable codes}
A STBC is said to be fast group decodable if it is multi-group decodable such that each group is fast decodable.
\begin{mydef} A STBC that encodes $2K$ real symbols is said to be FGD if its weight matrices are such that:
\begin{equation}
\begin{split}
\Am_k^{H}\Am_l+\Am_l^{H}\Am_k=\pmb{0},\ \forall \Am_k \in \Gc_i,\ \Am_l \in \Gc_j,\\
1\leq i\neq j\leq g,\ \vert\Gc_i\vert=n_i,\sum^{g}_{i=1}n_i=2K
\end{split}
\end{equation}
and that the weight matrices within each group are such that:
\begin{equation}
\begin{split}
\Am_k^{H}\Am_l+\Am_l^{H}\Am_k=\pmb{0},\ \forall \Am_k \in \Gc_{i,m},\ \Am_l \in \Gc_{i,n},\\
1\leq m\neq n\leq g_i,\ \vert\Gc_{i,j}\vert=n_{i,j},\ \sum^{g_i}_{j=1}n_{i,j}=k_i<n_i.
\end{split}
\end{equation}
where $\Gc_{i,m}$ (resp. $g_i$) denotes the set of weight matrices that constitute the $m$'th group  (resp. the number of inner groups) within the $i$'th group of symbols $\Gc_i$. 
\end{mydef} 
\noindent For instance, if a STBC that encodes $2K$ real symbols is FGD, its corresponding worst-case decoding complexity order for square QAM constellations is reduced from $M^K$ to $\sum^{g}_{i=1} \sqrt{M}^{n_i-k_i} \times \sum^{g_i}_{j=1}\sqrt{M}^{n_{i,j}-1}$. Similarly, if each group is conditionally orthogonal, the worst-case decoding complexity order is equal to $\sum^{g}_{i=1} \sqrt{M}^{n_i-k_i}$.
%************************************************************************************************************%
\begin{table*}[b!]
\hrule
\vspace*{2.4mm}
\setcounter{equation}{16}
\begin{equation}
\Xm_1(\sv)=\sqrt{\frac{2}{1+k^2}}\begin{bmatrix}
x_1+ikx_5&x_2+ikx_6&x_3+ikx_7&-ikx_4-x_8\\
-x_2+ikx_6&x_1-ikx_5&-ikx_4-x_8&-x_3-ikx_7\\
-x_3+ikx_7&ikx_4+x_8&x_1-ikx_5&x_2+ikx_6\\
ikx_4+x_8&x_3-ikx_7&-x_2+ikx_6&x_1+ikx_5 
\end{bmatrix} 
\label{newcode} 
\end{equation}
\end{table*}
\begin{table*}[t!]
\setcounter{equation}{19}
\begin{eqnarray}
x^{\text{ML}}_i\vert\left(\hat{x}_4,\hat{x}_9,\ldots,\hat{x}_{16}\right)&=&\textit{sign}\left(z_i\right)\times \text{min}\Big[\big\vert 2\ \textit{round}\big(\left(z_i-1\right)/2\big)+1\big\vert,\sqrt{M}-1\Big],\ i=1,2,3\\ 
x^{\text{ML}}_j\vert\left(\hat{x}_8,\hat{x}_9,\ldots,\hat{x}_{16}\right)&=&\textit{sign}\left(z_j\right)\times \text{min}\Big[\big\vert 2\ \textit{round}\big(\left(z_j-1\right)/2\big)+1\big\vert,\sqrt{M}-1\Big],\ j=5,6,7
\label{slicers}
\end{eqnarray}
\vspace*{2.4mm}
\hrule
\end{table*}
%******************************************************************************************************************%
\setcounter{equation}{15}
\section{The proposed FGD scheme}
Let the set $\left\lbrace\Id,\Rm_1,\ldots,\Rm_{2a+1}\right\rbrace$ denote the weight matrices of the square orthogonal STBC for $2^a$ transmit antennas \cite{CLIFF_ORTH}. The proofs of the following propositions are omitted due to space limitations.
\begin{myprop}
For $2^a$ transmit antennas, the two sets of matrices, namely $\Gc_1=\left\lbrace \Id,\Rm_1,\ldots,\Rm_a\right\rbrace\cup\Ac $ and $\Gc_2=\left\lbrace \Rm_{a+1},\ldots,\Rm_{2a+1}\right\rbrace \cup\Bc$ satisfy \eqref{g-group} where $\Ac$ and $\Bc$ are given in the Table \ref{casesI} and $\delta_{\Ac}(m)$ and $\delta_{\Bc}(m)$ are given in Table \ref{casesII}.
\end{myprop}
\begin{table}[h!]
\centering
\topcaption{Different cases for $\delta$}
\scriptsize
\begin{tabular}{|c|c|c|}
\hline
$a$&$\delta_{\Ac}(m)$&$\delta_{\Bc}(m)$\\
\hline
$4n$&$\frac{\left(\left(m\right)_4-1\right)\left(\left(m\right)_4-2\right)}{2}$&$\frac{2-\left(m\right)_4}{2}$\\
\hline
$4n+1$&$\frac{\left(\left(\left(m\right)_4\right)\left(\left(m\right)_4-1\right)\right)_4}{2}$&$\frac{\left(m\right)_4-1}{2}$\\
\hline
$4n+2$&$\frac{\left(\left(\left(m\right)_4\right)\left(\left(m\right)_4-3\right)\right)_4}{2}$&$\frac{\left(m\right)_4}{2}$\\
\hline
$4n+3$&$\frac{\left(\left(m\right)_4-2\right)\left(\left(m\right)_4-3\right)}{2}$&$\frac{3-\left(m\right)_4}{2}$\\
\hline
\end{tabular}
\label{casesII}
\end{table}
\begin{myprop}
The rate of the proposed family of FGD codes is equal to one complex symbol per channel use.
\end{myprop}
\noindent The weight matrices of our new FGD construction method for four and eight transmit antennas are listed in Table \ref{examples}.
\begin{table}[h!]
\centering
\topcaption{Examples of rate-1 FGD codes}
\scriptsize
\begin{tabular}{|c|c|c|}
\hline
Tx&$\Gc_1$&$\Gc_2$\\
\hline
4&$\Id,\Rm_2,\Rm_4,\Rm_1\Rm_3\Rm_5$&$\Rm_1,\Rm_3,\Rm_5,\Rm_2\Rm_4$\\
\hline
&$\Id,\Rm_2,\Rm_4,\Rm_6$&$\Rm_1,\Rm_3,\Rm_5,\Rm_7$\\
8&$j\Rm_1\Rm_3\Rm_5\Rm_7$&$j\Rm_2\Rm_4\Rm_6\Rm_1$\\
&$j\Rm_1\Rm_3\Rm_5\Rm_7\Rm_2$&$j\Rm_2\Rm_4\Rm_6\Rm_3$\\
&$j\Rm_1\Rm_3\Rm_5\Rm_7\Rm_4$&$j\Rm_2\Rm_4\Rm_6\Rm_5$\\
&$j\Rm_1\Rm_3\Rm_5\Rm_7\Rm_6$&$j\Rm_2\Rm_4\Rm_6\Rm_7$\\
\hline
\end{tabular}
\label{examples}
\end{table}

\subsection*{A new rate-1 FGD STBC for four transmit antennas}
According to Table \ref{examples}, the proposed rate-1 STBC in the case of four transmit antennas denoted $\Xm_1$ may be expressed as:
\begin{equation}
\begin{split}
\Xm_1(\sv)=&\Id x_1+\Rm_2 x_2+\Rm_4 x_3+\Rm_1\Rm_3\Rm_5 x_4+\\
&\Rm_1 x_5+\Rm_3 x_6+\Rm_5 x_7+\Rm_2\Rm_4 x_8.
\end{split}
\label{R1}
\end{equation}
According to \textbf{Definition 3}, the proposed code $\Xm$ is a FGD STBC with $g=2,n_1=n_2=4$ and $g_1=g_2=3$ such as $n_{i,j}=1,\ i=1,2,\ j=1,2,3$. Therefore, the worst-case decoding complexity order is $2\sqrt{M}$. However, the coding gain of $\Xm_1$ is equal to zero, in order to achieve the full-diversity, we resort to the constellation stretching \cite{CONST_STRETCHING} rather than the constellation rotation technique, otherwise the orthogonal symbols inside each group will be entangled together which in turns will destroy the FGD structure of the proposed code and causes a significant increase in the decoding complexity. 

The full diversity code matrix takes the form of \eqref{newcode} where $\sv=[x_1,\ldots,x_8]$ and $k$ is chosen to provide a high coding gain. The term $\sqrt{\frac{2}{1+k^2}}$ is added to normalize the average transmitted power per antenna per time slot.
\begin{myprop}
Taking $k=\sqrt{\frac{3}{5}}$, ensures the NVD property for the proposed code with a coding gain equal to 1.
\end{myprop}
%******************************************************************************************************************%
\setcounter{equation}{17}
\section{The proposed rate-2 code}
The proposed rate-2 code denoted $\Xm_2$ is simply obtained by multiplexing two rate-1 codes by means of a unitary matrix. Mathematically speaking, the rate-2 STBC is expressed as:
\begin{equation*}
\Xm_2\left(x_1,\ldots,x_{16}\right)=\Xm_1\left(x_1,\ldots,x_8\right)+e^{j\phi}\Xm_1\left(x_9,\ldots,x_{16} \right)\Um  
\end{equation*}
where $\Um$ and $\phi$ are chosen in order to maximize the coding gain. It was numerically verified for QPSK constellation that taking $\Um=j\Rm_1$ and $\phi_{\text{opt}}=\tan^{-1}\left(\frac{1}{2}\right)$ maximizes the coding gain which is equal to 1. To decode the proposed code, the receiver evaluates the QR decomposition of the real equivalent channel matrix $\tilde{\Hcb}$ \eqref{real_model}. The corresponding upper-triangular matrix $\Rm$ takes the form:
\begin{equation}
\Rm=\begin{bmatrix} \Am& \Bm\\ \zerov& \Cm \end{bmatrix}
\end{equation}
where $\Bm \in \mathbb{R}^{8 \times 8}$ has no special structure, $\Cm \in \mathbb{R}^{8 \times 8}$ is an upper triangular matrix and $\Am \in \mathbb{R}^{8 \times 8}$ takes the form:
\begin{equation}
\Am=\begin{bmatrix} x\ &\ 0\ &\ 0\ &\ x\ &\ 0\ &\ 0\ &\ 0\ &\ 0\\
                    0&x&0&x&0&0&0&0\\
                    0&0&x&x&0&0&0&0\\
                    0&0&0&x&0&0&0&0\\
                    0&0&0&0&x&0&0&x\\
                    0&0&0&0&0&x&0&x\\
                    0&0&0&0&0&0&x&x\\
                    0&0&0&0&0&0&0&x\end{bmatrix}
\end{equation}
in which $x$ indicates a possible non-zero position. For each value of $\left(x_9,\ldots,x_{16}\right)$, the decoder scans independently all possible values of $x_4$ and $x_8$, and assigns to them the corresponding 6 ML estimates of the rest of symbols via hard slicers according to (20)-\eqref{slicers}.
where:
\begin{eqnarray*}
z_i&=&\Big(y_i'-r_{i,4}\hat{x}_4-\sum^{16}_{k=9}r_{i,k}\hat{x}_k\Big)/r_{i,i},\ i=1,2,3\\
z_j&=&\Big(y_j'-r_{j,8}\hat{x}_8-\sum^{16}_{k=9}r_{j,k}\hat{x}_k\Big)/r_{j,j},\ j=5,6,7
\end{eqnarray*}
A rate-3/2 code that we will denote $\Xm_{3/2}$ may be easily obtained by puncturing the rate-2 proposed code $\Xm_2$ and may be expressed as:
\begin{equation*}
\Xm_{3/2}\left(x_1,\ldots,x_{12}\right)=\Xm_1\left(x_1,\ldots,x_8\right)+e^{j\phi_{\text{opt}}}\Xm_1\left(x_9,\ldots,x_{12} \right)\Um  
\end{equation*}
%*************************************************************************************************************%
\section{Numerical and simulation results}
In this section, we compare our proposed codes to comparable low-complexity STBCs existing in the literature in terms of worst-case decoding complexity, average decoding complexity and Bit Error Rate (BER) performance over quasi-static Rayleigh fading channels. One can notice from Table \ref{complexity} that the worst-case decoding complexity of the proposed rate-3/2 code is half that of the punctured rate-3/2 code in \cite{EX_CIOD} and is reduced by a factor of $\sqrt{M}/2$ w.r.t to the worst-case decoding complexity of the rate-3/2 code in \cite{4TX_MPLX_ORTH}. Moreover, the worst-case decoding complexity of our rate-2 code is half that of the code in \cite{EX_CIOD}. 
\begin{table}[h!]
\centering
\topcaption{summary of comparison in terms of worst-case decoding complexity}
\scriptsize
\begin{tabular}{|c|c|}
\hline
Code&Square QAM\\
&decoding complexity\\
\hline
The proposed rate-3/2 code&$2M^{2.5}$\\
\hline
The punctured P.Srinath-S.Rajan rate-3/2 code \cite{EX_CIOD}&$4M^{2.5}$\\  
\hline
The S.Sirianunpiboon \textit{et al.} code \cite{4TX_MPLX_ORTH}&$M^3$\\
\hline
The proposed rate-2 code&$2M^{4.5}$\\
\hline
The rate-2 P.Srinath-S.Rajan code \cite{EX_CIOD}&$4M^{4.5}$\\
\hline
\end{tabular}
\label{complexity}
\end{table}  
Simulations are carried out in a quasi-static Rayleigh fading channel in the presence of AWGN and 2 receive antennas for our rate-3/2 and rate-2 codes. The ML detection is performed via a depth-first tree traversal with infinite initial radius SD. The radius is updated whenever a leaf node is reached and sibling nodes are visited according to the simplified Schnorr-Euchner enumeration \cite{SE}. 
\begin{figure}[h!]
   \centering
     \includegraphics[scale=0.55]{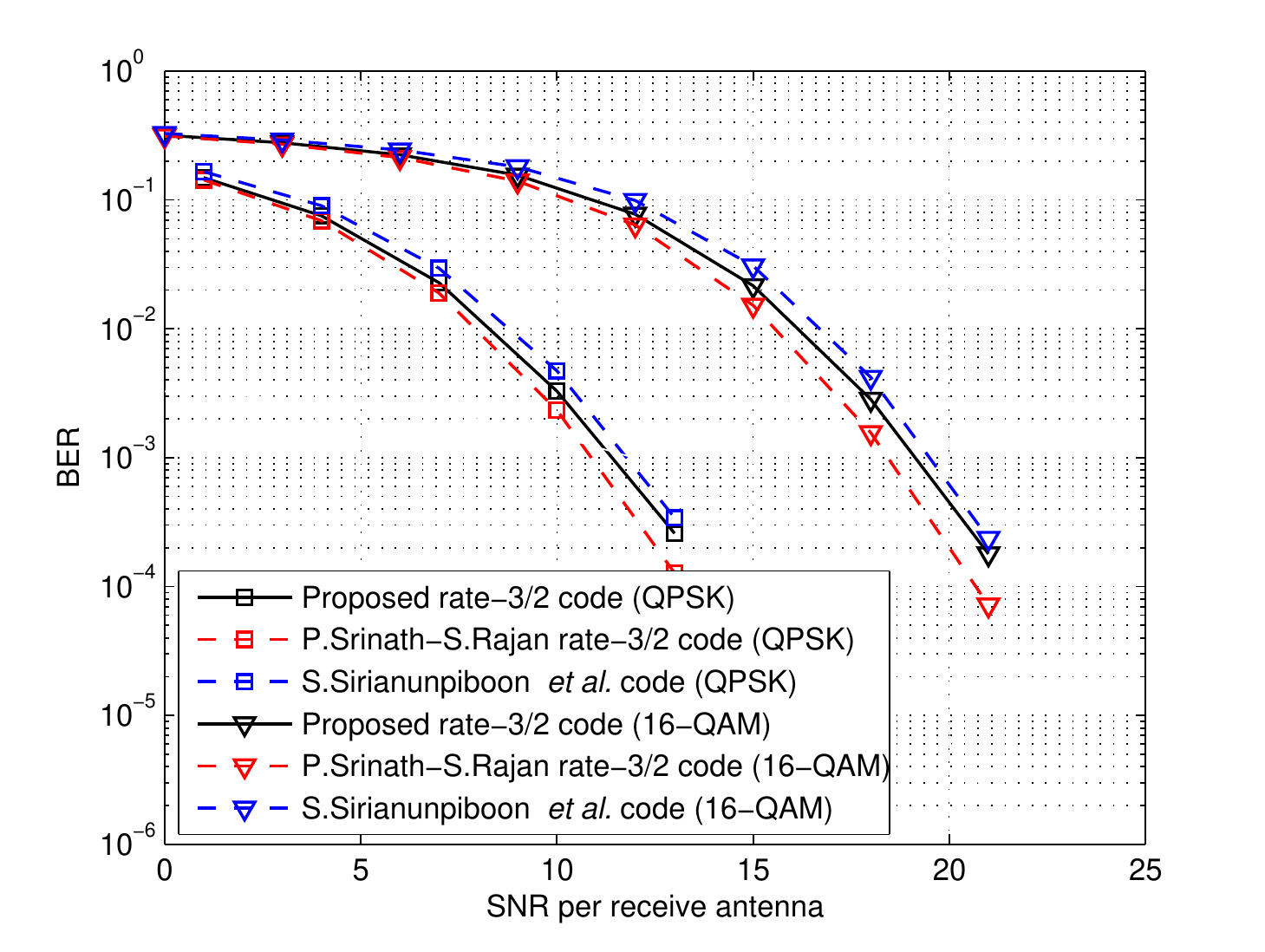} 
      \caption{BER performance for 4$\times$2 configuration}
      \label{BER}
\end{figure}
\begin{figure}[h!]
   \centering
     \includegraphics[scale=0.55]{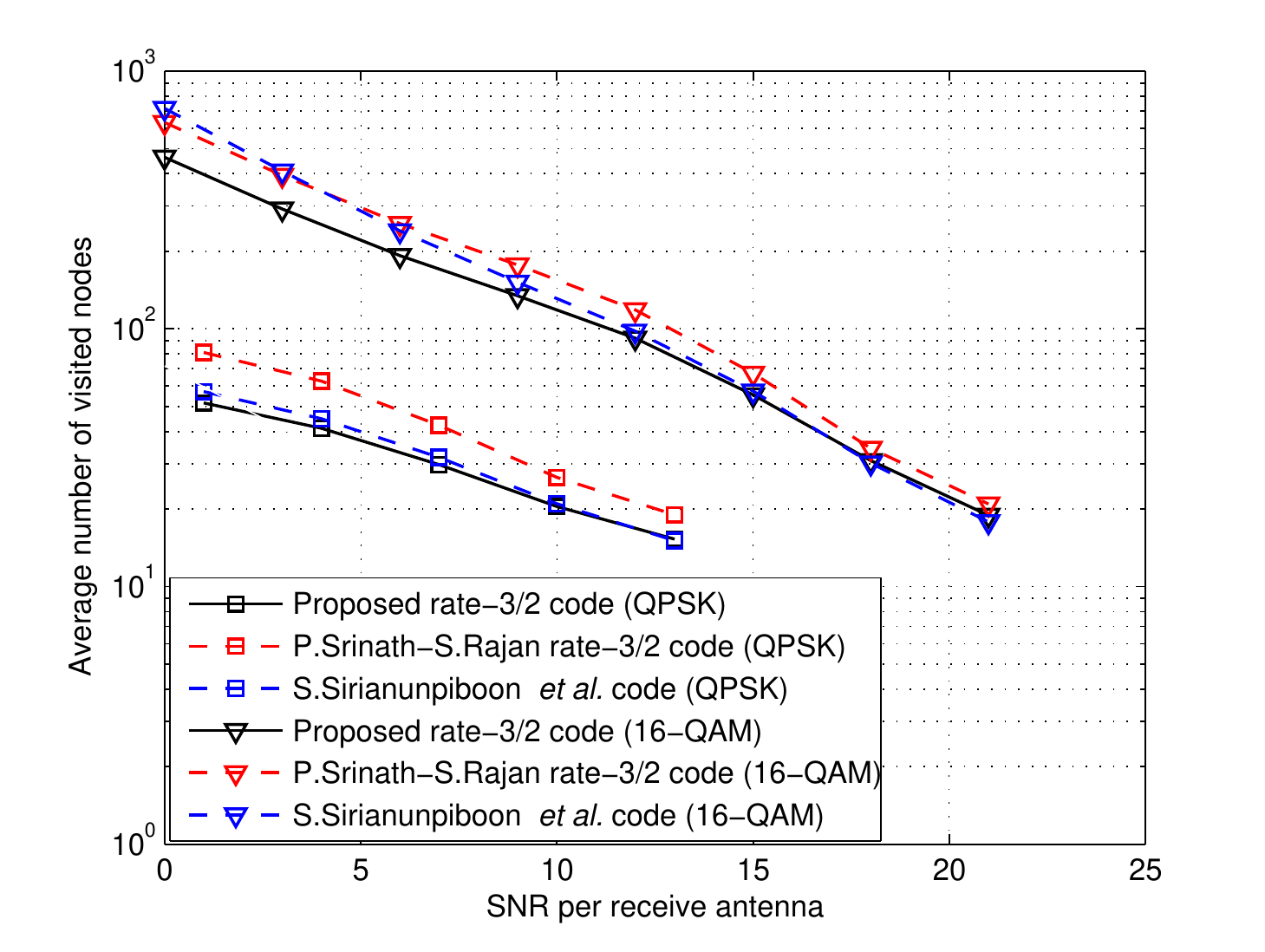} 
      \caption{Average complexity for 4$\times$2 configuration}
      \label{AVCOM}
\end{figure}
\begin{figure}[h!]
   \centering
     \includegraphics[scale=0.55]{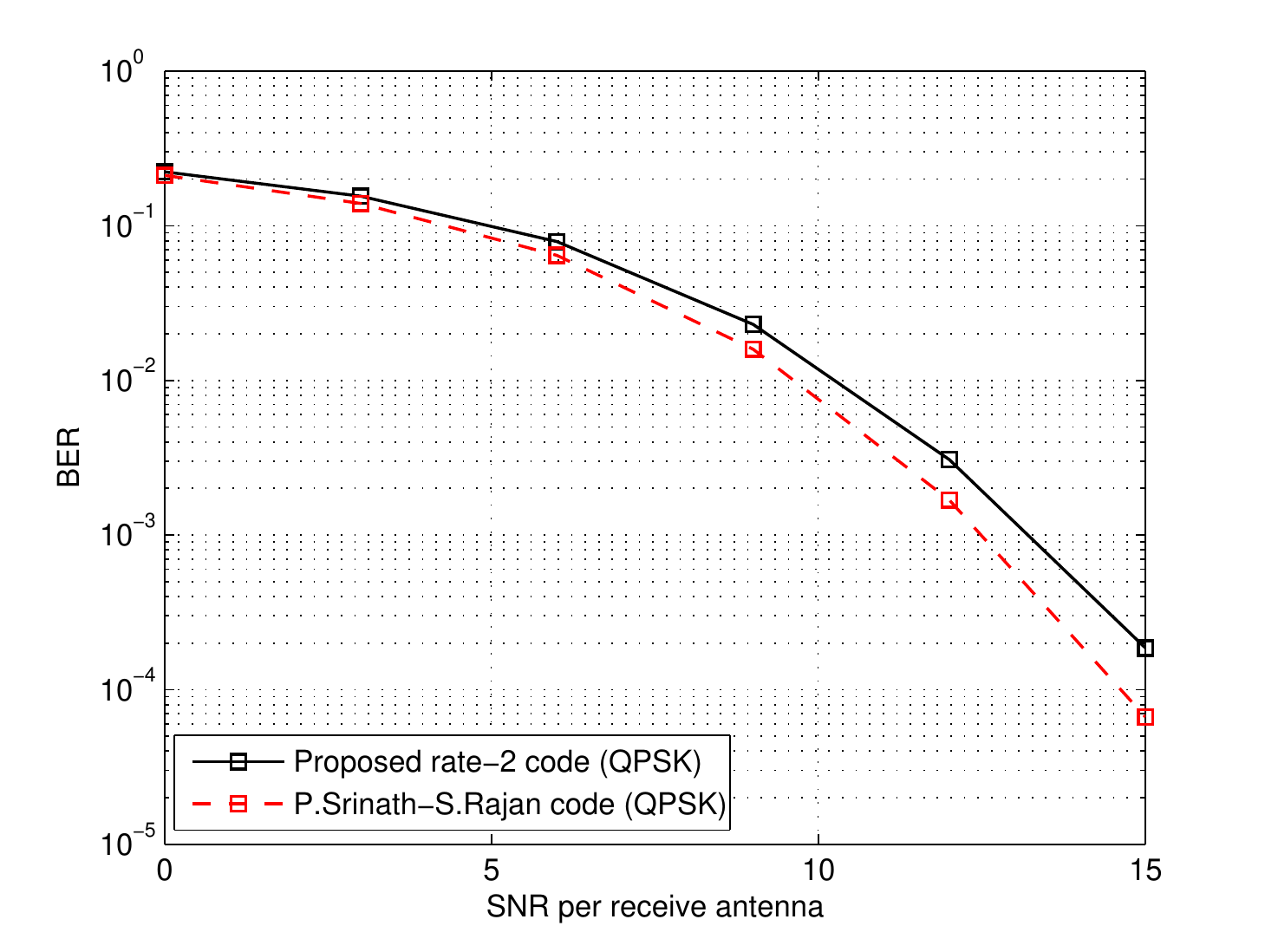} 
      \caption{BER performance for 4$\times$2 configuration}
      \label{BERQPSK}
\end{figure}
\begin{figure}[h!]
   \centering
     \includegraphics[scale=0.55]{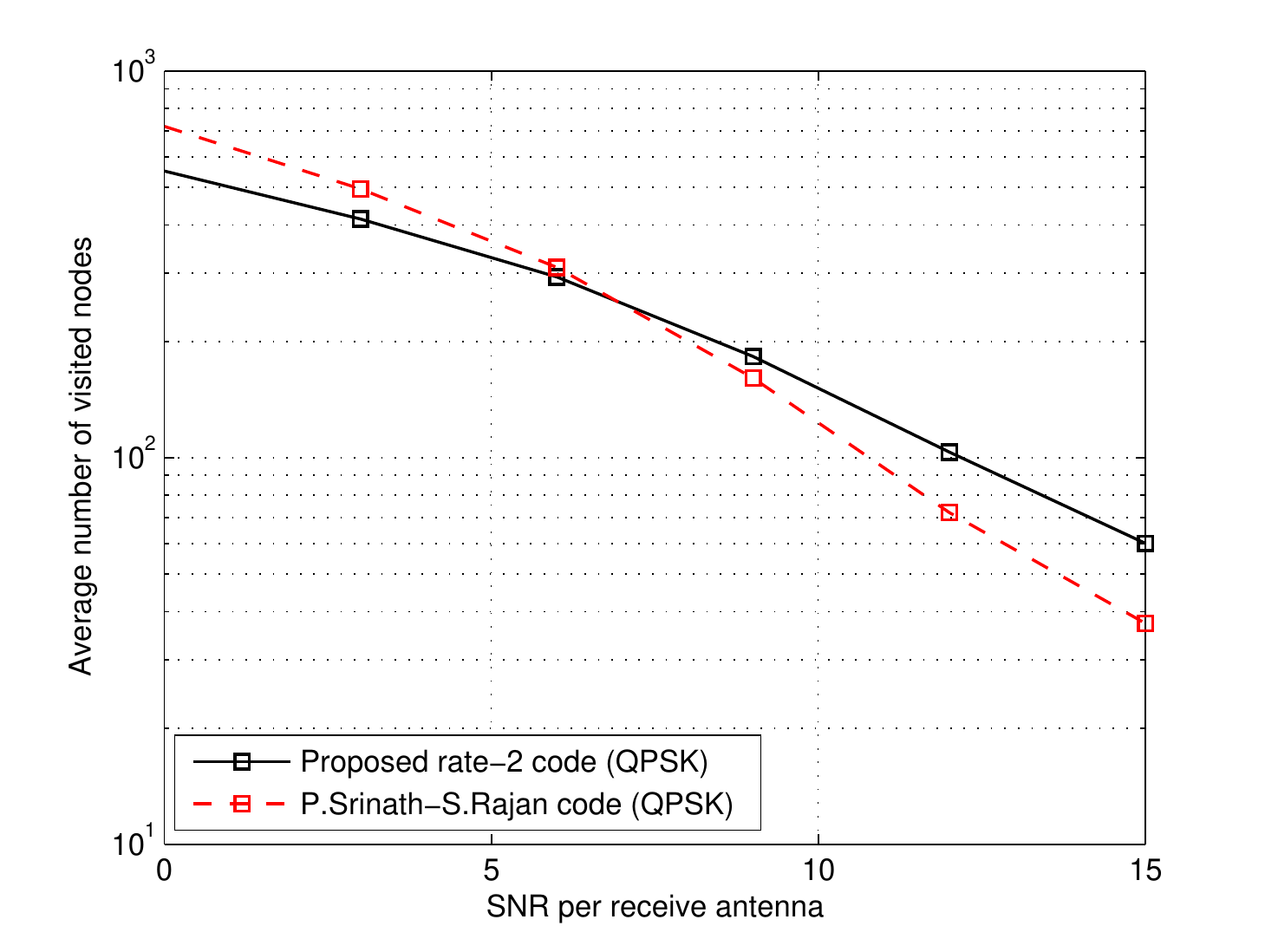} 
      \caption{Average Complexity for 4$\times$2 configuration}
      \label{AVCOMQPSK}
\end{figure}
From Fig.~\ref{BER}, one can notice that the proposed rate-3/2 code gains about 0.4 dB w.r.t to S.Sirianunpiboon \textit{et al.} code \cite{4TX_MPLX_ORTH} while it loses about 0.5 dB w.r.t to the punctured P.Srinath-S.Rajan code \cite{EX_CIOD} at $10^{-3}$ BER. From Fig. \ref{AVCOM}, it can be noticed that the average complexity of our rate-3/2 code is significantly less than that of the punctured P.Srinath-S.Rajan code \cite{EX_CIOD} and roughly equal to that of S.Sirianunpiboon \textit{et al.} code \cite{4TX_MPLX_ORTH}.
From Fig. \ref{BERQPSK}, one can notice that the proposed rate-2 code loses about 0.8 dB w.r.t the P.Srinath-S.Rajan code \cite{EX_CIOD} at $10^{-3}$ BER. However, from Fig. \ref{AVCOMQPSK} one can notice that our proposed code maintains its lower average decoding complexity in the low SNR region.
%*************************************************************************************************************%
\section{Conclusion}
In the present paper we have proposed a systematic approach for the construction of rate-1 FGD codes for an arbitrary number of transmit antennas. This approach when applied to the special case of four transmit antennas results in a new rate-1 FGD STBC that has the smallest worst-case decoding complexity among existing comparable low-complexity STBCs. The coding gain of the proposed FGD rate-1 code was then optimized through constellation stretching. Next we managed to increase the rate to 2 by multiplexing two rate-1 codes through a unitary matrix. A compromise between complexity and throughput may be achieved through puncturing the proposed rate-2 code which results in a new low-complexity rate-3/2 code. The worst-case decoding complexity of the proposed codes is lower than their STBC counterparts in the literature. Simulations results show that the proposed rate-3/2 code offers better performance that the S.Sirianunpiboon \textit{et al.} code \cite{4TX_MPLX_ORTH} but loses about 0.5 dB w.r.t the punctured P.Srinath-S.Rajan code \cite{EX_CIOD} at $10^{-3}$ BER. The proposed rate-2 code loses about 0.8 dB w.r.t the P.Srinath-S.Rajan code \cite{EX_CIOD} at $10^{-3}$ BER. In terms of average decoding complexity, we found that the proposed rate-3/2 code has a lower average decoding complexity while the proposed rate-2 code maintains its lower average decoding complexity in the low SNR region. 
%*************************************************************************************************************%

\end{document}